\documentclass[12pt]{article}
\usepackage[top=1.5cm, bottom=1.5cm, left=1.5cm, right=1.5cm]{geometry}
\usepackage{longtable}
\usepackage{natbib}
\usepackage{amsmath}
\usepackage{txfonts}
\usepackage{graphicx}
\usepackage{amssymb}
\usepackage[labelfont=bf]{caption}
\begin{document}
\title{The orbital period analyses for two cataclysmic variables: UZ Fornacis and V348 Puppis inside Period Gap}
\author{Z.-B. Dai $^{1,2}$, S.-B. Qian $^{1,2}$, E. Fern\'{a}ndez Laj\'{u}s $^{3}$ and G. L. Baume $^{3}$}

\footnotetext[1]{\scriptsize{National Astronomical Observatories/Yunnan Observatory,
Chinese Academy of Sciences, P. O. Box 110, 650011 Kunming, P. R.
China.}}

\footnotetext[2]{\scriptsize{United Laboratory of Optical Astronomy, Chinese Academy of
Science (ULOAC), 100012 Beijing, P. R. China.}}

\footnotetext[3]{\scriptsize{Facultad de Ciencias Astron\'omicas y Geof\'{\i}sicas, Universidad Nacional de La Plata and Instituto de Astrof\'{\i}sica de La Plata (CCT La Plata - CONICET/UNLP), Paseo del Bosque s/n, La Plata, Argentina.}}

\maketitle


\begin{abstract}
\small

\noindent{Four newest CCD eclipse timings of the white dwarf for polar UZ Fornacis and Six updated CCD mid-eclipse times for SW Sex type nova-like V348 Puppis are obtained. The detailed O-C analyses for both CVs inside period gap are made. Orbital period increases at a rate of $2.63(\pm0.58)\times10^{-11} s\;s^{-1}$ for UZ Fornacis and of $5.8(\pm1.9)\times10^{-12} s\;s^{-1}$ for V348 Puppis, respectively, are discovered in their new O-C diagrams. However, the conservative mass transfer from the secondary to the massive white dwarf cannot explain the observed orbital period increases for both CVs, which are regarded as part of modulations at longer periods. Moreover, the O-C diagram of UZ Fornacis shows a possible cyclical change with a period of $\sim23.4(\pm5.1)yr$. For explaining the observed cyclical period changes in UZ Fornacis, both mechanisms of magnetic activity cycles in the late-type secondary and the light travel-time effect are regarded as two probable causes. Not only does the modulation period 23.4yr obey the empirical correlation derived by \cite{lan99}, but also the estimated fractional period change $\Delta P/P\sim7.3\times10^{-7}$ displays a behavior similar to that of the CVs below the period gap. On the other hand, a calculation for the light travel-time effect implies that the tertiary component in UZ Fornacis may be a brown dwarf with a high confidence level, when the orbital inclination of the third body is larger than $16^{\circ}$.}

\end{abstract}


\begin{bfseries}
\noindent{Stars: cataclysmic variables; Stars: binaries : eclipsing; Stars: individual (UZ Fornacis); Stars : individual (V348 Puppis)}
\end{bfseries}




\section[]{Introduction}

A noted feature in binary period distribution of cataclysmic variables (CVs) is the existence a very few of CVs with orbital periods between 2 and 3 hours, which is the so-called "period gap" \citep{kat03}. Several theories are proposed for interpreting this phenomenon. One of the famous explanations \citep{rob81} is known as the interrupted-braking model, which indicates a sudden drop in the magnetic activity of the donor star, when it becomes full convective, results in its shrinkage within Roche lobe and temporarily cessation of mass transfer. Accordingly, based on CV period gap, a "standard" evolutionary theory suggests that a CV must continuously evolve towards shorter orbital period by losing angular momentum \cite[e.g.][]{spr83,rap83,kin88}.

UZ Fornacis was first detected in the EXOSAT archive (EXO 033319-2554.2) as a serendipitous X-ray source \citep{gio87}. The subsequent optical spectroscopies, polarimetries and photometries in many different brightness states \citep{ber88,ima98,bai91} established UZ Fornacis as an eclipsing polar (or AM Her type) with a high orbital inclination ($i\approx80^{\circ}$). The high-time-resolution optical photometry operated by \cite{per01} distinctly suggests that the strong magnetic field of the white dwarf channels the accretion flow onto both of the magnetic poles of the white dwarf. But based on the analysis of EUV and optical eclipse light curves \citep{,bai91,war95}, the impacting region (accretion spot) in lower hemisphere of the white dwarf is regarded as a major light source occulted by the mass donor star.

V348 Puppis was discovered as a faint X-ray source by the HEAO 1 satellite \citep{tuo90}. Following its spectroscopic classification, the multicolor photometries detected the V-shaped eclipses with large amplitude flickering at short wavelengths. The classification of V348 Puppis is disputed, since the X-ray data observed by \cite{tuo90,ros94} support a intermediate polar (or DQ Her subclass) candidate but optical spectra \citep{rod01,fro03} prefer a SW Sex type nova-like subclass identification. This means that V348 Puppis in period gap may be an important object for the study of its orbital period changes.

Since the period gap believed to be a transition zone in the evolutionary scenario, the stars in period gap can provide an opportunity to test the standard evolutionary theory of CVs. The orbital periods of UZ Fornacis and V348 Puppis, which are $\sim2.1$ hr and $\sim2.4$ hr respectively, definitely indicate that both objects are in period gap. Since the sharp eclipses in UZ Fornacis and V348 Puppis allow the determination of eclipse timings with high precision, the variations in the orbital periods can correctly reflect their secular evolutions. However, there is not any available orbital period analysis for V348 Puppis until now. As for UZ Fornacis, \cite{beu88} firstly detected a decrease of the orbital period, but the following orbital period analysis \citep{ram94,ima98} concluded that its orbital period is increasing, because they missed the four times of mid-eclipse obtained by \cite{ber88}. When the missed four data and the new data with high precision were added, \cite{per01} never measured any change in the orbital period for UZ Fornacis. Therefore, the rough orbital period analysis for both CVs in previous papers fails to deduce their possible evolutionary stages.

In this paper, The updated light curves near the mid-eclipse and the  eclipse timings for UZ Fornacis and V348 Puppis are presented in Sect. 2. Then Sect. 3 deals with the details of the O-C analysis for both CVs. Finally, the discussions of the possible mechanisms for orbital period changes are made in Sect. 4 and our principal conclusions in Sect. 5.

\section[]{Observations of eclipse timings}

\subsection{UZ Fornacis}

Four new times of light minimum are obtained from our CCD photometric observations with Roper Scientific, Versarray 1300B camera, with a thinned EEV CCD36-40 de 1340 $\times$ 1300 pix CCD chip, attached to the 2.15-m Jorge Sahade telescope at Complejo Astronomico El Leoncito (CASLEO), San Juan, Argentina. The BVR photometries were carried out on November 19, 2009. And the observation in I-filter was made on November 20, 2009. Two nearby stars which have the similar brightness in the same viewing field of telescope are chosen as the comparison star and the check star, respectively. All images were reduced by using PHOT (measure magnitudes for a list of stars) of the aperture photometry package of IRAF. The clock of the control computer operating the VersArray 1300B CCD camera is calibrated against UTC time by the GPS receiver's clock. The estimated precision of the time-stamp associated to each CCD frame is about 0.5s. An uniform exposure time for BVRI bands is adopted as $60s$, which is about the time resolution of each time-series $60.48s$. During our CASLEO observations, an estimate of the magnitude of UZ Fornacis is $V\sim18^{m}.1$, which implies that it is in a low brightness state at the time of our observations.

Inspections of Fig. 1 suggest that the eclipse depth of the white dwarf is clearly dependent on the band. Although the time resolution in our light curves is not high enough to identify the different occulted light source in detail, the eclipse light curve in I-filter shown in Fig. 2 may show the similar eclipse morphology as the descriptions in literatures \citep[e.g.][]{bai91,ima98}. The times of total eclipse and the flat-bottomed minima can be estimated as $t_{ad}\sim8$ min and $t_{bc}\sim4$ min. By using the midpoint times of steep ingress and egress of eclipse, we defined four mid-eclipse times of the primary accretion hot spot corresponding to BVRI bands, respectively. Considering the time-resolution of our observations is $\sim60$s, the quoted uncertainties are estimated as $0^{d}.0005$. The four new eclipse timings indicate that they are wavelength-dependent, with the eclipse occurring earlier at shorter wavelengths. This clearly implies that the composite eclipse of two sources, the bluer source being eclipsed earlier than the redder one. The average difference of the timings obtained in different wavelength is about $0^{d}.0003$. According to the eclipse phenomenology of UZ Fornacis \cite[e.g.][]{bai91,war95,ima98,per01}, there is a non-negligible uncertainty associated to the transformation of hot spot mid-eclipse timings into the white dwarf ones, which affects all timings except those of \cite{bai91}. Therefore, a common error bars of $0^{d}.00046$ are assigned to all data in the sample other than the timings obtained by \cite{bai91}. Moreover, in order to apply the uniform time system, the three eclipse timings in TDB observed by \cite{per01} and our new data in UTC are transformed to HJD. All of 44 times of light minimum from 1983 to 2009 for UZ Fornacis, including 4 new data from our observation, are listed in Table 1.

\subsection{V348 Puppis}

V348 Puppis was observed in 4 nights by using the 0.6m Helen Sawyer Hogg (HSH) telescope at Complejo Astronomico El Leoncito (CASLEO), San Juan, Argentina. The first three observations in 2008 November 28, 2009 January 14 and 16, were obtained with Photometric CH250, PM512 CCD camera. The final observation in 2009 November 24 was obtained with Apogee Alta U8300, Kodak KAF8300 chip. All CCD photometries were unfiltered. A nearby star which has the similar brightness in the same viewing field of telescope is chosen as the comparison star. All measurements have been corrected for differential extinction. At all 4 nights, the estimated magnitude of V348 Puppis is $V\sim15^{m}$, which is similar as the photometries \citep{rol00}. Moreover, the V-shaped eclipse of V348 Puppis is clearly shown in Fig. 3 as the literatures \citep[e.g.][]{rol00,rod01}. This imply that nova-like V348 Puppis may be a stable CVs, which is totally different from the polar UZ Fornacis. The calibrated clock used in observations is the same as that of UZ Fornacis. The exposure times for the observations in 2008, November 28, 2009, January 14, 16 and November 24 were 60s, 30s, 25s and 60s, respectively. Since the mid eclipse of V348 Puppis is V-shaped, which is different from that of UZ Fornacis, and six mid-eclipse timings were derived by using a parabolic fitting method to the deepest part of the eclipse. The uncertainties of the mid-eclipse timings in our observations are estimated to be $\sim0.^{d}0005$. Including the previous 48 times of light minimum for V348 Puppis \citep{tuo90,rol00}, we listed all 54 available times of light minimum covering about 21 yr in Table 2.

\section[]{Analyses of orbital period change}

\subsection{UZ Fornacis}

The updated linear ephemeris in HJD \citep{war95} is used to calculate the O-C values of 44 eclipse timings and, after linear revision, the new epochs and average orbital period of UZ Fornacis were derived as,
\begin{equation}
T_{min}=HJD\;2445567.175678(67)+0^{d}.0878654654(23)E,
\end{equation}
with standard deviation $\sigma_{1}=6^{d}.4\times10^{-4}$. The new calculated O-C values are listed in column 6 of Table 1. The new O-C diagram for UZ Fornacis shown in Fig. 4 implies a possible departure in the orbital period of UZ Fornacis. But a simple sinusoidal or quadratic fit cannot completely describe the variations in the upper panel of Fig. 4. Therefore,we attempted to use a quadratic-plus-sinusoidal ephemeris to fit the O-C values, and the least-square solution for O-C diagram of UZ Fornacis leads to
\begin{equation}
\begin{aligned}
(O-C)_{1} & =2^{d}.35(\pm0^{d}.65)\times10^{-3}-1^{d}.34(\pm0^{d}.31)\times10^{-7}E+1^{d}.16(\pm0^{d}.25)\times10^{-12}E^{2}\\
&+1^{d}.01(\pm0^{d}.34)\times10^{-3}\sin[0.0037^{\circ}(\pm0.0008^{\circ})E+283.9^{\circ}(\pm8.8^{\circ})],
\end{aligned}
\end{equation}
with standard deviation $\sigma_{2}=2^{d}.8\times10^{-4}$. Although the data near 17000 cycles present large scatters due to the early four data deriving from the light curves with very low time resolution \citep{ber88}, the reduced $\chi^{2}$ of Eq. 2 is calculated to be $\sim0.43$, which is lower than unity. Moreover, the data points shown in Fig. 4 clearly indicate that a cyclical period variation with a period of $23.4(\pm5.1)yr$ superimposed on a secular orbital period increase is significant. However, the small quadratic term in Eq. 2 implies that the quadratic ingredient in the O-C diagram of UZ Fornacis may be not significant. A linear-plus-sinusoidal ephemeris with the same modulation period as that in Eq. 2 is estimated to be
\begin{equation}
\begin{aligned}
(O-C)_{1} & =-4^{d}.8(\pm1^{d}.8)\times10^{-4}+5^{d}.9(\pm2^{d}.8)\times10^{-9}E\\
&+6^{d}.5(\pm1^{d}.2)\times10^{-4}\sin[0.0037^{\circ}(\pm0.0008^{\circ})E+52.8^{\circ}(\pm11.0^{\circ})],
\end{aligned}
\end{equation}
with standard deviation $\sigma_{3}=4^{d}.4\times10^{-4}$. Fig. 5 suggests that the best-fit sinusoid period for the Eq. 3 may be larger than 23 years. Thus, we allowed the sinusoidal period to be one of the free parameters of the fit. However, since the data points for UZ Fornacis only cover about a half of a whole sinusoid period, we cannot find a best-fit with a sinusoid period about twice the baseline of the current data. Accordingly, the further data points of UZ Fornacis are important for obtaining a precise linear-plus-sinusoidal ephemeris. In order to describe the significant level of the quadratic term in Eq. 2, a F-test as \cite{pri75} is attempted to apply for assessing the significance of quadratic terms in Eq. 2. The parameter $\lambda$ is described to be
\begin{equation}
\lambda=\frac{\sigma_{3}^{2}-\sigma_{2}^{2}}{\sigma_{2}^{2}/(n-6)},
\end{equation}
where $n$ is the number of data. Thus, a calculation gives $F(1,38)=55.8$, which suggests that it is significant well above $99.99\%$ level. According to Eq. 2, the orbital period increase rate of UZ Fornacis, $\dot{P}$, can be calculated to be $2.63(\pm0.58)\times10^{-11} s\;s^{-1}$, which is similar as the previous works \citep{ram94,ima98}.

\subsection{V348 Puppis}

The recent ephemeris derived by \cite{rod01} is used to calculate the O-C values of 54 eclipse timings for V348 Puppis. Since the investigation of Fig. 6 suggests larger scatters in observation season than the error of each data point, we assumed a common error bar $\sim0^{d}.0003$ for obtaining a fit with unity reduced $\chi^{2}$. The O-C diagram shown in Fig. 6 clearly implies an orbital period increase. Therefore, a quadratic ephemeris for V348 Puppis can be calculated to be
\begin{equation}
Min.I=2448591.66794(4)+0^{d}.101838934(6)E+2^{d}.94(\pm0^{d}.98)\times10^{-13}E^{2},
\end{equation}
with the reduced $\chi^{2}\sim1.2$. Although the large scatters in the O-C diagram prevent the further analysis for the orbital period changes in V348 Puppis at present, the observations over a longer baseline may be capable of detecting a cyclical variation with a large period. The orbital period increase rate of V348 Puppis, $\dot{P}$, can be estimated to be $5.8(\pm1.9)\times10^{-12} s\;s^{-1}$. Although the large scatters in observation seasons affect the uncertainty of $\dot{P}$, the average O-C diagram of V348 Puppis produced by averaging the mid-eclipse timings during each observation season significantly suggests the orbital period increase for V348 Puppis. Moreover, a F-test is used to assess the significance of quadratic term in Eq. 5. The parameter $\lambda=5.8$ indicates that it is significant with $\sim99\%$ level.

\section[]{Discussion}

\begin{figure}
\centering
\includegraphics[width=9.0cm]{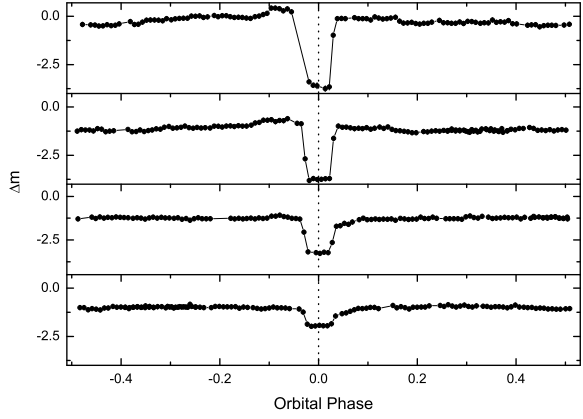}
\caption{The light curves of polar UZ Fornacis in BVRI bands are measured on 2009 November 19 and 20 by using the 2.15-m Jorge Sahade telescope at CASLEO. From top to bottom, the light curve are corresponding to B, V, R, I bands.} \label{Fig. 1}
\end{figure}

\begin{figure}
\centering
\includegraphics[width=9.0cm]{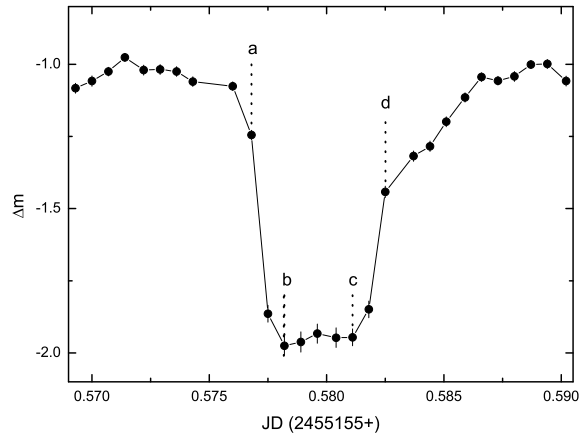}
\caption{The eclipse part of light curve of polar UZ Fornacis in I band is plotted. The a, b, c, d points roughly indicate the beginning and ending of ingress and egress, respectively. The approximate symmetrical profiles in the ingress and egress are clearly shown.} \label{Fig. 2}
\end{figure}

\begin{figure}
\centering
\includegraphics[width=9.0cm]{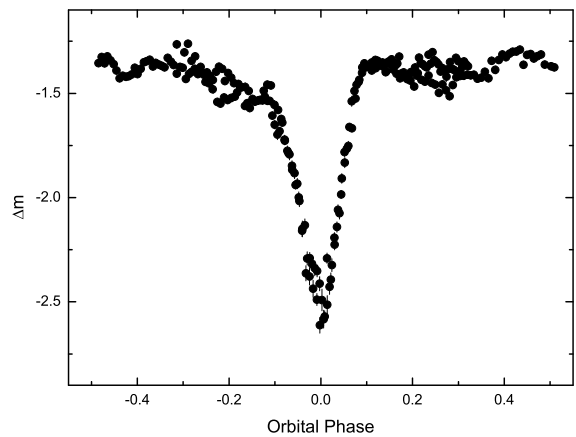}
\caption{The unfiltered light curve of V348 Puppis observed on 2008 November 28 by using the 0.6m Helen Sawyer Hogg (HSH) telescope at CASLEO is plotted. A V-shaped eclipse profile with probable time resolution is distinctly presented.} \label{Fig. 3}
\end{figure}

\begin{figure}
\centering
\includegraphics[width=9.0cm]{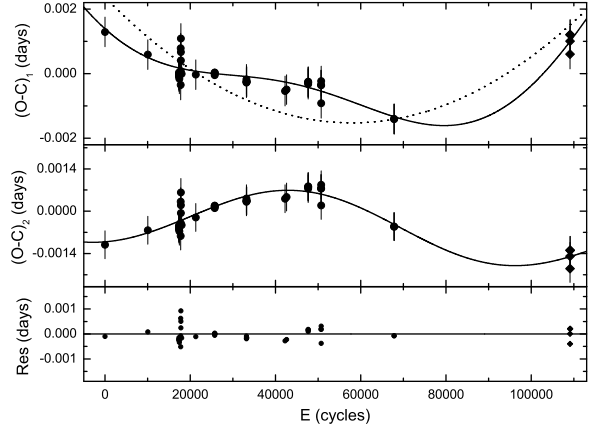}
\caption{The $(O-C)_{1}$ values of UZ Fornacis are fitted in the top panel with the ephemeris displayed in Eq. 2. After removing the quadratic element from the $(O-C)_{1}$ diagram, the $(O-C)_{2}$ values plotted in the middle panel significantly suggest a cyclical period change with a low amplitude. The residuals and their linear fitted solid line are presented in the bottom panel. The solid circles and diamonds in the three panels denote the data compiled from the previous papers and those we observed, respectively.} \label{Fig. 4}
\end{figure}

\begin{figure}
\centering
\includegraphics[width=9.0cm]{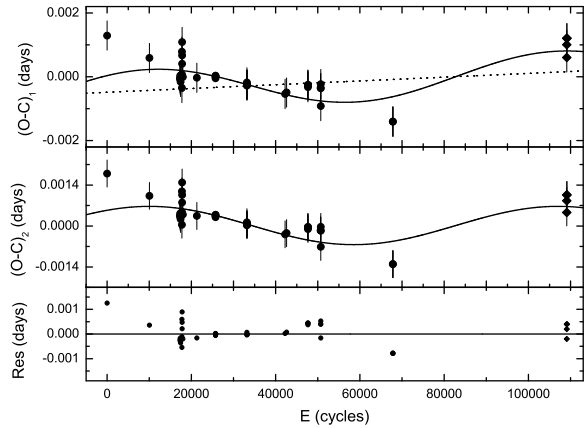}
\caption{The $(O-C)_{1}$ values of UZ Fornacis are fitted in the top panel with the ephemeris displayed in Eq. 3. After removing the linear element from the $(O-C)_{1}$ diagram, the middle panel significantly indicates a large departure from the fitted curve. The symbols are the same as that in Fig. 4.} \label{Fig. 4}
\end{figure}

\begin{figure}
\centering
\includegraphics[width=9.0cm]{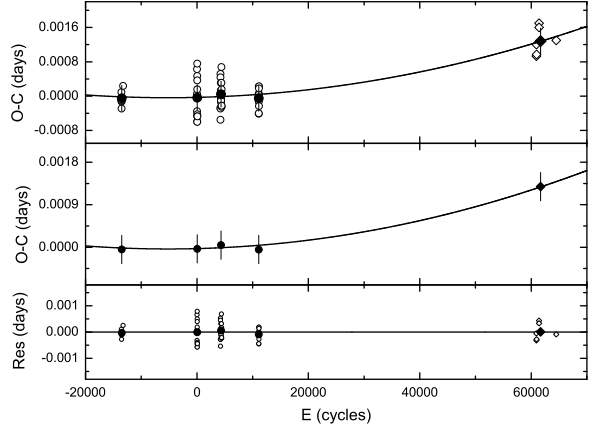}
\caption{The O-C diagrams of V348 Puppis are fitted by the quadratic ephemeris in Eq. 5. The open and solid symbols refer to the individual timings and the average ones, respectively. The residuals and their linear fitted solid line are presented in the bottom panel. The data in literatures and our observations are denoted by the cycles and diamonds, respectively.} \label{Fig. 5}
\end{figure}

\begin{figure}
\centering
\includegraphics[width=9.0cm]{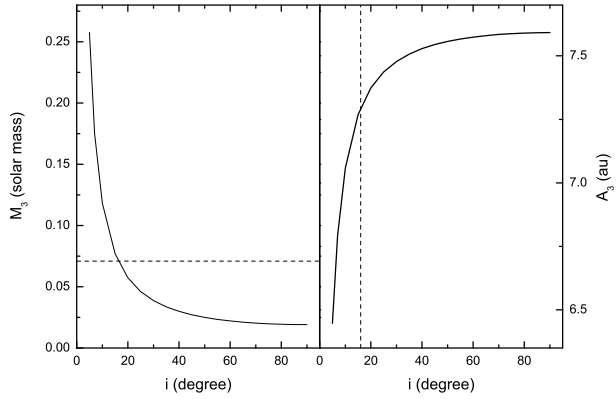}
\caption{The masses and separations of the third component in UZ Fornacis depending on its different orbital inclination $i$ are plotted in the left and right panels, respectively. The dash lines in the left and right panels denote a lower limit of mass of a star with hydrogen burning and the corresponding orbital inclination $i\sim16^{\circ}$, respectively.}
\label{Fig. 3}
\end{figure}

\subsection{Secular orbital period increase}

The normal mass transfer for both binaries from the red dwarf to the white dwarf are expected since the hot spots on the surface of the white dwarf and the occasional X-ray and UV flaring in UZ Fornacis are observed \citep[e.g.][]{bai91,war95,ima98,per01,pan02}, and V348 Puppis as a SW Sex-type system with an enhanced mass transfer rate are confirmed in \cite{rod01} and \cite{fro03}. Accordingly, the orbital period increase in both CVs should be predicted in principle. However, considering that the conservative mass transfer from a less massive star to a more massive one in a binary with the increasing orbital period and separation is sustainable if the mass-donor star expands faster than its Roche lobe, the conservative mass transfer for both CVs is not plausible explanation for the observed secular orbital period increase. The physical parameters of both CVs deduced by modeling the eclipse light curves for UZ Fornacis \citep{bai91,ima98}, and by measuring the radial velocity curves for V348 Puppis \citep{rod01}, respectively, indicate that both secondaries are the low-mass main sequence stars without violent expansion and the mass ratios of the secondary star to the white dwarf are less than unity. By using the system parameters of both CVs and the derived orbital period increase rates, we estimated the expansion rates of Roche lobe of the secondary stars, $\dot{R}_{rl2}$, caused by mass transfer in the case of conservation to be $\sim1.1\times10^{-8}R_{\odot}yr^{-1}$ and $\sim2.5\times10^{-9}R_{\odot}yr^{-1}$ for UZ Fornacis and V348 Puppis, respectively. Moreover, the calculated conservative mass transfers for both CVs may be adiabatic mass transfers since the mass loss timescales for both secondaries are lower than their Kelvin-Helmholtz timescales. According to the results calculated in polytropic models, a low-mass main sequence star as mass-donor star in a binary with a rapid mass transfer (i.e. adiabatic mass transfer) may expand as a exponent $\sim1/3$ \citep{hje87}. Thus, the possible expansion rates of the mass-donor stars $\dot{R}_{2}$ can be estimated to be $2.9\times10^{-9}R_{\odot}yr^{-1}$ and $7.8\times10^{-10}R_{\odot}yr^{-1}$ for UZ Fornacis and V348 Puppis, respectively, which are a factor of $\sim3$ less than the respective $\dot{R}_{rl2}$. This paradox may imply that both CVs in period gap are interesting objects and worthy the further observations for investigating their secular orbital period variations. On the other hand, since the current data with large scatters for both CVs limit the precise detection of secular orbital period increase, the possibility that the observed increase trend in the O-C diagram may reflect part of a variation another modulation in a longer timescale cannot be totally ruled out.

\subsection{A cyclical period variation in UZ Fornacis}

A notable cyclical period variation with a period of $\sim23.4(\pm5.1)yr$ is presented in the middle panel of Fig. 4 after removing the quadratic element. However, the only one cycle of the modulation and the large scatters shown in Fig. 4 suggest that it is yet not possible to judge if the modulation is strictly periodic or not. Since the spectral type of the secondary of UZ Fornacis is regarded as M4-M5 dwarf \citep{fer89,pan02}, a solar-type magnetic activity cycle in its convective shell should be considered for explaining the observed oscillations in the O-C diagram \citep{hal89,app92,lan06}. According to the regression relationship between the length of the modulation period $P_{mod}$ and the angular velocity $\Omega$ derived by \cite{lan99}
\begin{equation}
Log P_{mod}=-0.36(\pm0.10)Log \Omega+0.018,
\end{equation}
a slop $\sim-0.43$ for UZ Fornacis can be estimated. This suggests that UZ Fornacis can fit the relationship well. Furthermore, according to Eq. 2, the fractional period change $\Delta P/P$ for UZ Fornacis can be estimated to be $\sim7.3\times10^{-7}$. Since the orbital period of UZ Fornacis is about $2.1hr$ which is inside period gap, the $\Delta P/P$ value of UZ Fornacis can offer a good sample for checking the diagram of $\Delta P/P$ versus $\Omega$ of the active component star for CVs \citep{bor08}. A calculation indicates that UZ Fornacis obviously deviates from the relation $\Delta P/P \propto \Omega^{-0.7}$ and displays a behavior similar to that of the CVs below the period gap \citep{bor08}. However, considering that the generation of large scale magnetic fields in fully convective stars may be different from the stars with radiative cores \citep{dob05}, magnetic activity cycles still may be a possible explanation for UZ Fornacis. On the other hand, using an extended model with Applegate's considerations introduced by \cite{lan05,lan06}, the longest timescale of energy dissipation $\tau\approx0.8yr$ for the magnetic active cycles in UZ Fornacis can be estimated from the eigenvalue of the equation of angular momentum conservation $\lambda_{20}\approx3.95\times10^{-8}s^{-1}$. Moreover, the scaling relationship of the dissipated power and the luminosity of the secondary stars derived by \cite{lan06} suggests that the needed energy cannot be sustained by stellar luminosity of UZ Fornacis. This means that a reconsideration of the hypothesis of magnetic active cycles is needed to interpret the observed cyclical period variations in CVs.

Another plausible mechanism for the cyclical period variation is the light travel-time effect, which is caused by a perturbation from a tertiary component. An inspection of Fig. 4 may suggest that the orbital eccentricity of the third star is near zero. By using the amplitude of sinusoidal curve and the Third Kepler Law, the projected distance $a^{'}\sin(i)$ from binary to the mass center of the triple system and the mass function of the third component $f(m_{3})$ can be calculated to be $0.17(\pm0.06)au$ and $9.4(\pm0.9)\times10^{-6}M_{\odot}$, respectively. We set a combined mass of $0.7M_{\odot}+0.14M_{\odot}$ for the eclipsing pair of UZ Fornacis. When the orbital inclination of the third component in this system is large than $16^{\circ}$, it may be a brown dwarf with a confidence level $\geq82\%$ based on the inclination. According to the both relationships described in Fig. 7, the mass of the third star in UZ Fornacis is close to a lower limit of brown dwarf's mass $\sim0.014M_{\odot}$ \citep{cha00,bur01} as long as the inclination of the third star is higher than $85^{\circ}$ and the distance from the third body to the mass center of system is $\sim7.6 au$, which is over three orders of magnitude larger than the separation of binary estimated from the orbital period of UZ Fornacis. Thus, this third star can survive the previous common envelope evolution of the parent binary. Moreover, the mass of this third star implies that it may be a critical substellar object between brown dwarf and giant planet.

\section{Conclusion}

The orbital period variations in two CVs with deep eclipse inside period gap, UZ Fornacis and V348 Puppis are investigated in detail. The new eclipse timings, comprising 4 data points for polar UZ Fornacis and 6 data points for V348 Puppis, respectively, suggest that the orbital period increase with a rate of $2.63(\pm0.58)\times10^{-11} s\;s^{-1}$ for UZ Fornacis and $5.8(\pm1.9)\times10^{-12} s\;s^{-1}$ for V348 Puppis, respectively, are observed in the O-C diagrams. However, based on the physical parameters of both CVs, the observed secular orbital period increase shown in Fig. 4 and Fig. 6 cannot be explained by the conservative mass transfer since $\dot{R}_{rl2}>\dot{R}_{2}$ in the case of conservation for both CVs. Thus, the increasing trend shown in the O-C diagrams for both CVs may be just part of another modulation at a longer period. Moreover, the failure of a linear-plus-sinusoidal ephemeris for UZ Fornacis may imply that current data in O-C diagram only cover half-cycle modulation.

Aside of the observed orbital period increase in both CVs, a cyclical period variation with a low amplitude of $1^{d}.01(\pm0^{d}.34)\times10^{-3}$ and a period of $23.4(\pm5.1)yr$ in UZ Fornacis is discovered. Both mechanisms of magnetic activity cycles and light-travel effect apparently can interpret the observed O-C oscillation. As for the former, the modulation period fits the regression relationship derived by \cite{lan99}, and the derived fractional period change $\Delta P/P\sim7.3\times10^{-7}$ displays a behavior similar to that of the CVs below the period gap in the diagram of $\Delta P/P$ versus $\Omega$ \citep{bor08}. Moreover, since the longest timescale of energy dissipation $\tau\approx0.8yr$ estimated by an extended model with Applegate's hypothesis introduced by \cite{lan05,lan06} does not support Applegate's mechanism for interpreting the observed modulation in UZ Fornacis, a new available model of magnetic active cycles is expected. On the other hand, a calculation about light travel-time effect indicates that a brown dwarf as the third component in a $0.7M_{\odot}+0.14M_{\odot}$ eclipsing pair of UZ Fornacis is possible with a confidence level $\geq82\%$. Assuming the inclination of the tertiary component is high, the third star may be a critical substellar object between brown dwarf and giant planet. It is important to note that uncertainties in the system parameters of both binaries affect the precision of the detailed O-C analysis for both CVs in period gap, and extending the observational base line for checking all the hypotheses discussed above is needed for a more comprehensive investigation of their orbital period changes.

\section*{Acknowledgments}

This work was partly Supported by Special Foundation of President of The Chinese Academy of Sciences and West Light Foundation of The Chinese Academy of Sciences, Yunnan Natural Science Foundation (2008CD157), the Yunnan Natural Science Foundation (No. 2005A0059M) and Chinese Natural Science Foundation (No.10573032, No. 10573013 and No.10433030). CCD photometric observations of UZ Fornacis and V348 Puppis were obtained with the 0.6-m Helen Sawyer Hogg telescope and 2.15-m Jorge Sahade telescope at CASLEO, San Juan, Argentina. The Apogee Alta U8300 CCD system used has been provided by the Instituto de Astronom\'{\i}ay  F\'{\i}sica del Espacio, Conicet/Universidad Nacional de Buenos Aires, Argentina. We thank the referee very much for the helpful comments and suggestions that helped to improve this paper greatly.

\begin{center}
\begin{longtable}{p{2.5cm}cccccc}
\caption{The 44 eclipse timings for the polar UZ Fornacis.}\\
\hline\hline
\hspace{2em}JD.Hel. & type & error(d) & Method & E (cycle) & $(O-C)^{d}$ & Ref.\\
\hspace{1.5em}2400000+&&&&&&\\
\hline
\endfirsthead
\caption{Continued.}\\
\hline\hline
\hspace{2em}JD.Hel. & type & error(d) & Method & E (cycle) & $(O-C)^{d}$ & Ref.\\
\hspace{1.5em}2400000+&&&&&&\\
\hline
\endhead
\hline
\endfoot
\hline
\multicolumn{7}{p{12cm}}{\scriptsize{Note. $^{*}$ the unknown state. $^{q}$ the low state. $^{f}$ the faint high state. $^{i}$ the intermediate state. $^{h}$ the high state.  $^{BVRI}$ the observed bands. References: (1) \cite{osb88}; (2) \cite{beu88}; (3) \cite{fer89}; (4) \cite{ber88}; (5) \cite{ram94}; (6) \cite{bai91}; (7) \cite{war95}; (8) \cite{ima98}; (9) \cite{per01}; (10) This paper.}}
\endlastfoot
45567.1768700$^{*}$ & pri &  .00046 & EXOSAT & 0      &  .00129 & (1)\\
46446.9730700$^{*}$ & pri &  .00046 & EXOSAT & 10013  &  .00059 & (1)\\
47088.7418100$^{q}$ & pri &  .00046 & ccd    & 17317  & -.00003 & (2)\\
47089.7082700$^{q}$ & pri &  .00046 & ccd    & 17328  & -.00009 & (2)\\
47090.5870500$^{q}$ & pri &  .00046 & ccd    & 17338  &  .00003 & (2)\\
47091.5535000$^{q}$ & pri &  .00046 & ccd    & 17349  & -.00004 & (2)\\
47094.7166200$^{q}$ & pri &  .00046 & ccd    & 17385  & -.00007 & (2)\\
47097.7918200$^{q}$ & pri &  .00046 & ccd    & 17420  & -.00016 & (2)\\
47127.1387000$^{f}$ & pri &  .00046 & ccd    & 17754  & -.00035 & (3)\\
47127.2270000$^{f}$ & pri &  .00046 & ccd    & 17755  &  .00008 & (3)\\
47127.7549000$^{q}$ & pri &  .00046 & pe     & 17761  &  .00079 & (4)\\
47127.8431000$^{q}$ & pri &  .00046 & pe     & 17762  &  .00109 & (4)\\
47128.7213000$^{q}$ & pri &  .00046 & pe     & 17772  &  .00067 & (4)\\
47128.8089000$^{q}$ & pri &  .00046 & pe     & 17773  &  .00041 & (4)\\
47145.0636000$^{f}$ & pri &  .00046 & ccd    & 17958  & -.00001 & (3)\\
47437.9191700$^{*}$ & pri &  .00046 & pe     & 21291  & -.00003 & (5)\\
47827.9541300$^{q}$ & pri &  .00006 & ccd    & 25730  &  .00002 & (6)\\
47828.0419900$^{q}$ & pri &  .00006 & ccd    & 25731  &  .00002 & (6)\\
47828.1298700$^{q}$ & pri &  .00006 & ccd    & 25732  &  .00003 & (6)\\
47829.0084400$^{q}$ & pri &  .00006 & ccd    & 25742  & -.00005 & (6)\\
47829.0963500$^{q}$ & pri &  .00006 & ccd    & 25743  & -.00001 & (6)\\
47829.1842100$^{q}$ & pri &  .00006 & ccd    & 25744  & -.00001 & (6)\\
48482.7271658$^{*}$ & pri &  .00046 & ROSAT  & 33182  & -.00028 & (5)\\
48482.9029958$^{*}$ & pri &  .00046 & ROSAT  & 33184  & -.00018 & (5)\\
48483.3422558$^{*}$ & pri &  .00046 & ROSAT  & 33189  & -.00025 & (5)\\
48483.4301258$^{*}$ & pri &  .00046 & ROSAT  & 33190  & -.00025 & (5)\\
48483.6058458$^{*}$ & pri &  .00046 & ROSAT  & 33192  & -.00026 & (5)\\
49276.6792538$^{q}$ & pri &  .00046 & EUVE   & 42218  & -.00054 & (7)\\
49310.3317728$^{q}$ & pri &  .00046 & EUVE   & 42601  & -.00049 & (7)\\
49752.6467558$^{h}$ & pri &  .00046 & pe     & 47635  & -.00027 & (8)\\
49753.6132158$^{h}$ & pri &  .00046 & pe     & 47646  & -.00033 & (8)\\
49755.5463358$^{h}$ & pri &  .00046 & pe     & 47668  & -.00025 & (8)\\
49755.6341658$^{h}$ & pri &  .00046 & pe     & 47669  & -.00028 & (8)\\
50018.7032958$^{f}$ & pri &  .00046 & pe     & 50663  & -.00035 & (8)\\
50020.7236358$^{f}$ & pri &  .00046 & pe     & 50686  & -.00092 & (8)\\
50021.6908358$^{f}$ & pri &  .00046 & pe     & 50697  & -.00023 & (8)\\
50021.7785758$^{f}$ & pri &  .00046 & pe     & 50698  & -.00036 & (8)\\
51522.4319264$^{i}$ & pri &  .00046 & STJ    & 67777  & -.00140 & (9)\\
51528.4067702$^{i}$ & pri &  .00046 & STJ    & 67845  & -.00141 & (9)\\
51528.4946341$^{i}$ & pri &  .00046 & STJ    & 67846  & -.00141 & (9)\\
55154.6168380$^{Vq}$ & pri & .00046 & ccd    & 109115 &  .00101 & (10)\\
55154.7048690$^{Rq}$ & pri & .00046 & ccd    & 109116 &  .00121 & (10)\\
55154.7922168$^{Bq}$ & pri & .00046 & ccd    & 109117 &  .00061 & (10)\\
55155.5835680$^{Iq}$ & pri & .00046 & ccd    & 109126 &  .00121 & (10)\\
\end{longtable}
\end{center}


\begin{center}
\begin{longtable}{p{2.5cm}cccccc}
\caption{The 54 eclipse timings for the SW Sex type nova-like V348 Puppis.}\\
\hline\hline
\hspace{2em}JD.Hel. & type & error(d) & Method & E (cycle) & $(O-C)^{d}$ & Ref.\\
\hspace{1.5em}2400000+&&&&&&\\
\hline
\endfirsthead
\caption{Continued.}\\
\hline\hline
\hspace{2em}JD.Hel. & type & error(d) & Method & E (cycle) & $(O-C)^{d}$ & Ref.\\
\hspace{1.5em}2400000+&&&&&&\\
\hline
\endhead
\hline
\endfoot
\hline
\multicolumn{7}{p{14cm}}{\scriptsize{References: (1) \cite{tuo90}; (2) \cite{rol00}; (3) This paper.}}
\endlastfoot
47210.121125 & pri & .0003 & pe  & -13566 &  .00009 & (1)\\
47211.954000 & pri & .0003 & pe  & -13548 & -.00013 & (1)\\
47212.055875 & pri & .0003 & pe  & -13547 & -.00010 & (1)\\
47212.157708 & pri & .0003 & pe  & -13546 & -.00010 & (1)\\
47213.990625 & pri & .0003 & pe  & -13528 & -.00029 & (1)\\
47240.061917 & pri & .0003 & pe  & -13272 &  .00024 & (1)\\
48591.769462 & pri & .0003 & ccd & 1      & -.00035 & (2)\\
48592.686379 & pri & .0003 & ccd & 10     &  .00002 & (2)\\
48593.705382 & pri & .0003 & ccd & 20     &  .00063 & (2)\\
48593.806642 & pri & .0003 & ccd & 21     &  .00006 & (2)\\
48594.722715 & pri & .0003 & ccd & 30     & -.00042 & (2)\\
48594.824510 & pri & .0003 & ccd & 31     & -.00047 & (2)\\
48595.640158 & pri & .0003 & ccd & 39     &  .00047 & (2)\\
48595.742287 & pri & .0003 & ccd & 40     &  .00076 & (2)\\
48596.657488 & pri & .0003 & ccd & 49     & -.00059 & (2)\\
48596.759320 & pri & .0003 & ccd & 50     & -.00060 & (2)\\
48597.676629 & pri & .0003 & ccd & 59     &  .00016 & (2)\\
48597.778677 & pri & .0003 & ccd & 60     &  .00037 & (2)\\
48598.694383 & pri & .0003 & ccd & 69     & -.00047 & (2)\\
48598.796652 & pri & .0003 & ccd & 70     & -.00004 & (2)\\
49020.613001 & pri & .0003 & ccd & 4212   & -.00055 & (2)\\
49020.715111 & pri & .0003 & ccd & 4213   & -.00027 & (2)\\
49021.632451 & pri & .0003 & ccd & 4222   &  .00052 & (2)\\
49021.734218 & pri & .0003 & ccd & 4223   &  .00044 & (2)\\
49022.650034 & pri & .0003 & ccd & 4232   & -.00029 & (2)\\
49022.752227 & pri & .0003 & ccd & 4233   &  .00006 & (2)\\
49027.640565 & pri & .0003 & ccd & 4281   &  .00013 & (2)\\
49029.779356 & pri & .0003 & ccd & 4302   &  .00031 & (2)\\
49030.695522 & pri & .0003 & ccd & 4311   & -.00008 & (2)\\
49030.797304 & pri & .0003 & ccd & 4312   & -.00014 & (2)\\
49031.612099 & pri & .0003 & ccd & 4320   & -.00005 & (2)\\
49031.714122 & pri & .0003 & ccd & 4321   &  .00013 & (2)\\
49032.630438 & pri & .0003 & ccd & 4330   & -.00010 & (2)\\
49032.732446 & pri & .0003 & ccd & 4331   &  .00007 & (2)\\
49039.657734 & pri & .0003 & ccd & 4399   &  .00031 & (2)\\
49039.759941 & pri & .0003 & ccd & 4400   &  .00068 & (2)\\
49040.573722 & pri & .0003 & ccd & 4408   & -.00025 & (2)\\
49040.675596 & pri & .0003 & ccd & 4409   & -.00022 & (2)\\
49041.592531 & pri & .0003 & ccd & 4418   &  .00016 & (2)\\
49720.653953 & pri & .0003 & ccd & 11086  & -.00041 & (2)\\
49720.756257 & pri & .0003 & ccd & 11087  &  .00006 & (2)\\
49721.571129 & pri & .0003 & ccd & 11095  &  .00022 & (2)\\
49721.672783 & pri & .0003 & ccd & 11096  &  .00004 & (2)\\
49722.589171 & pri & .0003 & ccd & 11105  & -.00013 & (2)\\
49722.690736 & pri & .0003 & ccd & 11106  & -.00040 & (2)\\
49723.607919 & pri & .0003 & ccd & 11115  &  .00023 & (2)\\
49723.709706 & pri & .0003 & ccd & 11116  &  .00018 & (2)\\
49723.811139 & pri & .0003 & ccd & 11117  & -.00023 & (2)\\
54798.649900 & pri & .0003 & ccd & 60949  &  .00093 & (3)\\
54798.752050 & pri & .0003 & ccd & 60950  &  .00120 & (3)\\
54800.686720 & pri & .0003 & ccd & 60969  &  .00097 & (3)\\
54845.598410 & pri & .0003 & ccd & 61410  &  .00170 & (3)\\
54847.635140 & pri & .0003 & ccd & 61430  &  .00160 & (3)\\
55159.771100 & pri & .0003 & ccd & 64495  &  .00130 & (3)\\
\end{longtable}
\end{center}

\end{document}